\documentclass[aps,prl,showpacs,twocolumn]{revtex4}
\usepackage{bm,color,amsmath,amssymb,mathrsfs,latexsym,graphicx,psfrag}

\newcommand{\bs}[1]{\boldsymbol{#1}}

\newcommand{\ket}[1]{\left|#1\right\rangle}

\newcommand{\of}[1]{\!\left(#1\right)}

\newcommand{\expect}[1]{\left\langle#1\right\rangle}

\def\ie{{\it i.e.},\ }
\def\eg{{\it e.g.}\ }
\def\ea{{\it et al.}}
\def\eV{}

\begin{document}
\title{No evidence for spontaneous orbital currents in finite size
  studies\\ of three-band models for CuO planes} \author{Martin
  Greiter and Ronny Thomale} \affiliation{Institut f\"ur Theorie der
  Kondensierten Materie, Universit\"at Karlsruhe, D 76128 Karlsruhe}
\pagestyle{plain}

%\homepage[]{Your web page}
%\thanks{}
%\altaffiliation{}
\date{\today}

\begin{abstract}
  We have numerically evaluated the current-current correlations for
  three-band models of the CuO planes in high-$T_{\rm c}$
  superconductors at hole doping $x=1/8$.  The results show no
  evidence for the orbital current patterns proposed by Varma.  If
  such patterns exist, the associated energy is estimated to be
  smaller than 5 meV per link even if
  $\epsilon_\text{p}-\epsilon_\text{d}=0$.  Assuming that the
  three-band models are adequate, quantum critical fluctuations of
  such patterns hence cannot be responsible for phenomena occurring at
  significantly higher energies, such as superconductivity or the
  anomalous properties of the pseudogap phase.
\end{abstract}

% 74.20.Mn  Nonconventional mechanisms (spin fluctuations, polarons 
%           and bipolarons, resonating valence bond model, anyon mechanism, 
%           marginal Fermi liquid, Luttinger liquid, etc.)
% 74.72.-h  Cuprate superconductors (high-Tc and insulating parent compounds)
% 74.20.-z  Theories and models of superconducting state
% insert suggested keywords - APS authors don't need to do this
%\keywords{}

\maketitle

The problem of high-$T_{\rm c}$ superconductivity has been something
like a holy grail to the field of condensed matter physics for the
past two decades~\cite{zaanen-06np138}.  In the words of R.B.\
Laughlin, it has not been a fight, but a war.  It has been a traumatic
experience for some of those involved, but has also led to a plethora
of new developments extending far beyond the field.  Many of the
experimental techniques used to study the systems, like angle resolved
photo emission spectroscopy (ARPES) or scanning tunneling microscopy
(STM), have undergone revolutions with regard to resolution and data
processing.  {The} theory of superconductivity in high-$T_{\rm c}$
cuprates has been found as many times as victories must have been
proclaimed in civil wars, but while individuals believe to have the
theory, there is no consensus what the theory should be.  Many ideas,
even though too general to qualify as complete theories of the
cuprates, have inspired a vast amount of research in both high-$T_{\rm
  c}$ and other areas.  Most prominently among them are the notions of
a resonating valence bond (RVB) state~\cite{Anderson87s1196}, the
gauge theories of antiferromagnetism~\cite{Lee-06rmp17}, and the
notion of quantum criticality~\cite{Sachdev03rmp913}.  There have
been, however, a few concise proposals which make falsifiable
predictions.  Intellectual masterpieces among them have been the
theory of anyon superconductivity~\cite{Laughlin88s525}, the proposal
of kinetic energy savings through interlayer
tunneling~\cite{Anderson95s1154}, the SO(5) theory of a common order
parameter for superconductivity and magnetism~\cite{Demler-04rmp909},
and a more recent proposal that the anomalous properties of the
cuprates may be due to quantum critical fluctuations of current
patterns formed spontaneously in the CuO planes~\cite{Varma99prl3538,
  Varma06prb155113}.  This last proposal is further investigated in
this Letter.

The idea of a spontaneous symmetry breaking through orbital currents
was, as with so many major advances in physics, motivated by
experiment.  The normal state of the cuprates at optimal doping shows
a behavior which can be classified as quantum critical, and has been
rather adequately described by a phenomenological theory called
marginal Fermi liquid~\cite{Varma-89prl1996}.  
This phenomenology suggests a quantum critical point (QCP) at a hole
doping level of $x_{\rm c}\approx 0.19$, an assumption consistent with
a significant body of experimental data~\cite{Tallon-01pc53,
  Alff-03n698, vanderMarel03n271, dagan-04prl167001,
  Naqib-05prb054502}.  Critical fluctuations around this point would
then be responsible for the anomalous properties of the pseudogap
phase, and provide the pairing force responsible for the
superconducting phase which hides the QCP.

Interpreting the phase diagram in these terms, one is immediately led
to ask what the phase to the left of the QCP, \ie for $x<x_{\rm c}$,
might be.  The theory would require a spontaneously broken symmetry
beyond the global U(1) symmetry broken through superconductivity
(which is often erroneously refered to as a broken gauge
symmetry~\cite{Greiter05ap217}).  In addition, as the fluctuations are
assumed to determine the phase diagram up to temperatures of several
hundred Kelvin, the characteristic energy scale of the correlations
inducing this symmetry violation must be at least of the same order of
magnitude.  No definitive evidence of such a broken symmetry has been
found up to now, even though several possibilities have been suggested.
These include stripes~\cite{Kivelson-03rmp1201}, a
$d$-density wave~\cite{Chakravarty-01prb094503}, and most recently a
checkerboard charge density wave~\cite{Li-06prb184515}.

The general consensus is that the low energy sector of the three-band
Hubbard model proposed for the CuO planes (see \eqref{3bH}
below)~\cite{emery87prl2794} reduces to a one-band $t$--$t'$--$J$
model, with parameters $t\approx 0.44$, $t'\approx -0.06$, and
$J\approx 0.128$ (energies throughout this article are in
eV)~\cite{Zhang-88prb3759, Eskes-88prl1415, Hybertsen-89prb9028,
  Hybertsen-90prb11068, Rice-91ptrsla459}.  For the undoped CuO
planes, the formal valances are Cu$^{2+}$ and O$^{2-}$.  As the
electron configuration of Cu atoms is [Ar] $3d^{10}4s^1$, this implies
one hole per unit cell, which will predominantly occupy the
$3d_{x^2-y^2}$ orbital.  As the onsite potential $\epsilon_\text{p}$
in the O $2p_x$ and $2p_y$ orbitals relative to the Cu $3d_{x^2-y^2}$
orbital is generally assumed to be of the order of
$\epsilon_\text{p}=3.6\eV$ (with $\epsilon_{\text{d}}=0$), and hence
smaller than the onsite Coulomb repulsion $U_d\approx 10.5\eV$ for a
second hole in the $3d_{x^2-y^2}$ orbital, it is clear that additional
holes doped into the planes will primarily reside on the Oxygens.  The
maximal gain in hybridization energy is achieved by placing the
additional hole in a combination of the surrounding O $2p_x$ and
$2p_y$ orbitals with the same symmetry as the original hole in the Cu
$3d_{x^2-y^2}$ orbital, which requires antisymmetry of the wave
function in spin space, \ie the two holes must form a singlet.  This
picture is strongly supported by data from NMR~\cite{Walstedt90s248}
and even more directly from spin-resolved
photoemission~\cite{tjeng-97prl1126}.  In the effective one-band
$t$--$J$ model description of the CuO planes, these singlets
constitute the ``holes'' moving in a background of spin 1/2 particles
localized at the Cu sites.

\begin{figure}[t]
  \begin{minipage}[c]{0.3\linewidth}
    \includegraphics[width=\linewidth]{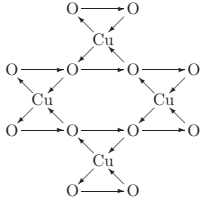}
  \end{minipage}
\caption{Orbital current pattern proposed by Varma.}
\label{fig:pattern}
\vspace{-4mm}
\end{figure}
In contrast to this picture, Varma~\cite{Varma99prl3538,
  Varma06prb155113} has proposed that the additional holes doped in
the CuO planes do not hybridize into Zhang-Rice singlets, but give
rise to circular currents on O-Cu-O triangles, which align into a
planar pattern as shown in Fig.~\ref{fig:pattern}.  He assumes that
the inter-atomic Coulomb potential $V_\text{pd}$ is larger than both
the hopping $t_\text{pd}$ and the onsite potential $\epsilon_\text{p}$
of the O $2p$ orbitals relative to the Cu $3d_{x^2-y^2}$ orbitals, an
assumption which is not consistent with the values generally agreed on
(see the list below \eqref{3bH}).
Making additional assumptions, Varma has shown that the circular
current patterns are stabilized in a mean field solution of the
three-band Hubbard model.  The orbital current patterns break
time-reversal symmetry breaking (T) and the discrete four-fold
rotation symmetry on the lattice, but leave translational symmetry
intact.  The current pattern is assumed to disappear at a doping level
of about $x_{\rm c}\approx 0.19$.  The phenomenology of CuO
superconductors, including the pseudogap and the marginal Fermi liquid
phase, are assumed to result from critical fluctuations around this
QCP, as outlined above.

Motivated by this proposal, several experimental groups have looked
for signatures of orbital currents or T violation in CuO
superconductors.  While there is no agreement between different groups
regarding the manifestation of T violation in ARPES 
studies~\cite{Kaminski-02n610,Borisenko-04prl207001}, a recent
neutron scattering experiment by Fauqu\'e \ea~\cite{fauque-06prl197001} 
indicates magnetic order within the unit cells of the CuO planes.
Their results are consistent with Varma's proposal, and call the
validity of the one-band models into question.  

In a recent article, Aji and Varma~\cite{Aji-06cm0610646} have mapped
the four possible directions of the current patterns in each unit cell
onto two Ising spins, and investigated the critical fluctuations.
Within this framework, the coupling between and the transverse fields
for these Ising spins decide whether or under which circumstances the
model displays long-range order in the orbital currents.  

We hence undertook to estimate these couplings through numerical
studies of finite clusters containing 8 unit cells, \ie 8 Cu and 16 O
sites, and periodic boundary conditions (which do not frustrate but
should enhance the correlations).  The total number of holes on our
cluster was taken $N=9$ (5 up-spins and 4 down-spins), corresponding
to a hole doping of $x=1/8$.  We had hoped that the energy associated
with a domain wall, which may be implemented through a twist in the
boundary conditions, would provide information regarding the coupling
aligning the orbital currents in neighboring plaquets, while the
splitting between the lowest energies for a finite system would
provide an estimate for the transverse field.
The result of our endeavors, however, is a daunting disappointment:
The coupling is zero within the error bars of our numerical
experiments.

Let us now report our numerical studies in detail.  To begin with, we
wish to study the three-band Hubbard Hamiltonian
$H=H_\text{t}+H_\text{U}$ with
\begin{eqnarray}
  H_\text{t}&\!=\!&\! 
  \sum_{i,\sigma} \epsilon_\text{p}\, 
  n_{i,\sigma}^{\text{p}}
  -t_{\text{pd}}\!\sum_{\langle i,j \rangle , \sigma}\! 
  \left(d_{i,\sigma}^{\dagger} p^{\phantom{\dagger}}_{j,\sigma}
    +p_{j,\sigma}^{\dagger}d_{i,\sigma}^{\phantom{\dagger}}\right)
  \nonumber \\
  &-\!&\! t_{\text{pp}}\!\sum_{\langle i,j \rangle , \sigma}\!
  \left( p_{i,\sigma}^{\dagger}p_{j,\sigma}^{\phantom{\dagger}}
    +p_{j,\sigma}^{\dagger}p_{i,\sigma}^{\phantom{\dagger}}\right)
  + V_{\text{pd}}\hspace{-9pt}\sum_{\langle i,j \rangle , \sigma, \sigma '}
  \hspace{-9pt} n_{i, \sigma}^{\text{d}} n_{j,\sigma '}^{\text{p}},\hspace{3pt}  
  \nonumber \\[2pt]
  H_\text{U}\!\!&\!=\!&\! 
  U_{\text{p}}\sum_i n_{i, \uparrow}^{\text{p}} n_{i, \downarrow}^{\text{p}}
  + U_{\text{d}}\sum_i n_{i, \uparrow}^{\text{d}} n_{i, \downarrow}^{\text{d}},
  \label{3bH}
\end{eqnarray} 
where {\small $\langle\ ,\ \rangle$} indicates that the sums extend over pairs
of nearest neighbors, while $d_{i,\sigma}$ and $p_{j,\sigma}$
annihilate holes in Cu $3d_{x^2-y^2}$ or O $2p$ orbitals, respectively.
Hybertsen \ea~\cite{Hybertsen-89prb9028} assumed $t_{\text{pd}}=1.5\eV$,
$t_{\text{pp}}=0.65\eV$, $U_{\text{d}}=10.5\eV$, $U_{\text{p}}=4\eV$,
$V_{\text{pd}}=1.2\eV$, and $\epsilon_{\text{p}}=3.6\eV$.

\begin{figure}[t]
  \begin{minipage}[c]{0.9\linewidth}
    \includegraphics[width=\linewidth]{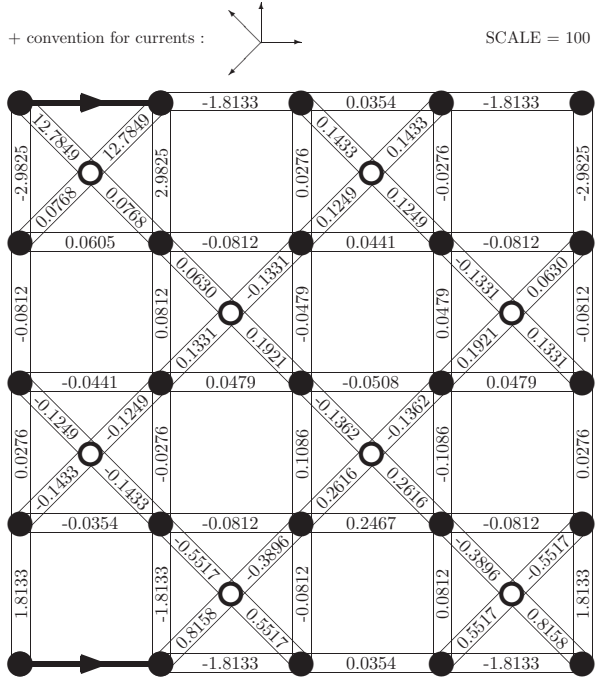}
  \end{minipage}
  \caption{Current-current correlations $\expect{j_{k,k+\hat
        x}j_{l,m}}$ multiplied by $10^2$ for the ground state of
    \eqref{3btj} with $\epsilon_\text{p}=3.6$ on a 24 site cluster (8
    Cu = open circles, 16 O = filled circles) with PBCs.  The
    reference link is indicated in the top and (due to the PBCs)
    bottom left corner.}
\label{fig:ep36}
\vspace{-3mm}
\end{figure} 

In order to be able to diagonalize \eqref{3bH} for 24 sites, we need
to truncate the Hilbert space.  A first step is to eliminate doubly
occupied sites, which yields the effective three-band $t$--$J$
Hamiltonian
\begin{eqnarray}
  H_{\text{eff}}\!\!&=\!\!&\tilde{P}_{\text{G}}  H_\text{t}
  \tilde{P}_{\text{G}} + H_\text{J} \ \ \ \text{with}\\[5pt]
  H_\text{J}&\!\!=\!\!&
  J_{\text{pd}}\hspace{-1pt}\sum_{\langle i, j \rangle} 
  \of{\bs{S}_{i}^{\text{p}}\!\cdot\! \bs{S}_{j}^{\text{d}}-\frac{1}{4}}
  + J_{\text{pp}}\hspace{-1pt}\sum_{\langle i, j \rangle} 
  \of{\bs{S}_{i}^{\text{p}}\!\cdot\! \bs{S}_{j}^{\text{p}}-\frac{1}{4}},
  \hspace{5pt}\nonumber
  \label{3btj}
\vspace{-5pt}
\end{eqnarray}
where
%\begin{equation}
%  \label{jpdjpp}
%  J_{\text{pd}}=2t_{\text{pd}}^2\of{\frac{1}{U_{\text{d}}-\epsilon_\text{p}}
%    +\frac{1}{U_\text{p}+\epsilon_\text{p}}}\ \text{and}\
%  J_{\text{pp}}=\frac{4t_\text{pp}^2}{U_\text{p}}.
%  \nonumber
%\end{equation}
\begin{equation}
  \label{jpdjpp}
  J_{\text{pd}}=2t_{\text{pd}}^2\of{\frac{1}{U_{\text{d}}-\epsilon_\text{p}}
    +\frac{1}{U_\text{p}+\epsilon_\text{p}}}, \
  J_{\text{pp}}=\frac{4t_\text{pp}^2}{U_\text{p}}\hspace{1pt},
  \nonumber
\end{equation}
and the sums in $H_\text{J}$ are limited to pairs where both neighbors
are occupied by holes.  If $\tilde{P}_{\text{G}}$ only eliminates
configurations with more than one hole on a site, the dimension of the
$S^z_\text{tot}=\frac{1}{2}$ sub-sector is with 164,745,504 still
beyond our capabilities.  We have hence implemented two further ways
of truncating the Hilbert space: (a) We limit the maximal number of
holes allowed in the O orbitals to $N^\text{max}_\text{ox}$.  (b) We
limit the maximal number of CuO links occupied with 2 holes to
$N^\text{max}_\text{link}$.  For the values proposed by Hybertsen \ea,
either truncation should hardly affect the low energy sector.  In
Table~\ref{tab:one},
\begin{table}[b]
  \centering
  \begin{tabular}{ccc||ccc}
    \hline\hline
    $N^\text{max}_\text{ox}$   & $E$ & $10^2\!\expect{JJ}_\text{diag.}$ &
    $N^\text{max}_\text{link}$ & $E$ & $10^2\!\expect{JJ}_\text{diag.}$ \\
\hline
    3& -0.705  & -0.0085   & 2 & -0.063  &  0.0004   \\
    4& -0.835  & -0.0304   & 3 & -0.635  & -0.0279   \\
    5& -0.877  & -0.0488   & 4 & -0.851  & -0.0508   \\
    \hline\hline
  \end{tabular}
  \caption{Ground state energies per unit cell and a current-current
    correlation for various truncations of the Hilbert space.}
  \label{tab:one}
\end{table}
we compare the ground state energies and the current-current
correlation on diagonally and maximally separated links on the torus
for different values of $N^\text{max}_\text{ox}$ and
$N^\text{max}_\text{link}$.  This comparison gives us confidence that
it is not unreasonable to set $N^\text{max}_\text{link}=4$, which
reduces the Hilbert space dimension to 93,595,824.  Since the
truncation (b) is predominantly projecting out states with high
kinetic energy, we expect that the value of $\epsilon_{\text{p}}$ will
not affect the validity of the approximation.  Excluding
configurations with too many CuO links occupied by two holes should,
in any occasion, not weaken the tendency to form orbital current
patterns.  Truncation (a), by contrast, is no longer reasonable if
$\epsilon_\text{p}$ is small.

Let us now turn to our results for the current-current correlations in
the ground state (situated at the M point in the Brillouin zone) for
our 24 site cluster with 9 holes and parameters as assumed by
Hybertsen, where no orbital current pattern are expected.  With the
current operator for, \eg an O-O link given by
\begin{equation}
  \label{curdef}
  j_{k,l}=i t_{\text{pp}}\sum_\sigma
  \of{p_{l,\sigma}^{\dagger}p_{k,\sigma}^{\phantom{\dagger}}
     -p_{k,\sigma}^{\dagger}p_{l,\sigma}^{\phantom{\dagger}}},
\vspace{-3pt}
\end{equation}
the correlations function $\expect{j_{k,k+\hat x}j_{l,m}}$ with an O-O link as
reference link is depicted in Fig.~\ref{fig:ep36}.  As expected, the
correlations fall off rapidly, and there is no indication of order.

\begin{figure}[t]
  \begin{minipage}[c]{0.9\linewidth}
    \includegraphics[width=\linewidth]{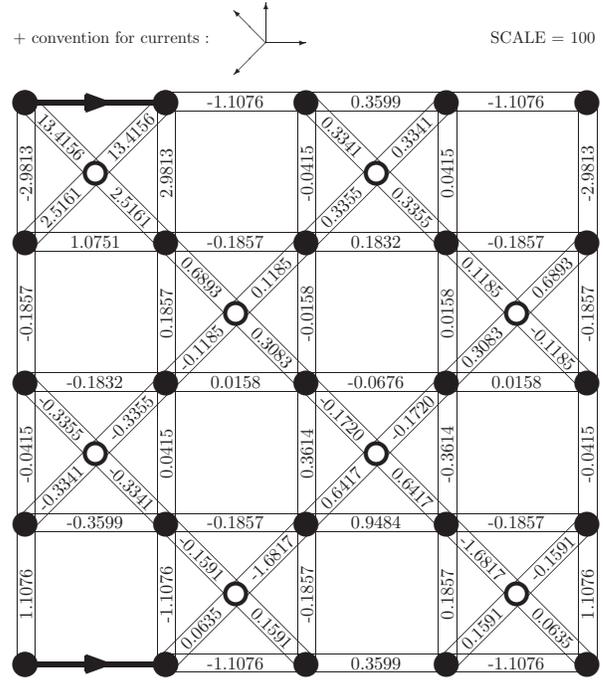}
  \end{minipage}
  \caption{as in Fig.~\ref{fig:ep36}, but with
    $\epsilon_{\text{p}}=0$.  Except for the vertical links, positive
    correlations indicate alignment with the current pattern shown in
    Fig.~\ref{fig:pattern}.}
\label{fig:ep0}
\vspace{-3mm}
\end{figure}

The crucial result is that the correlations change continuously and
only quantitatively, but not really qualitatively, as
$\epsilon_{\text{p}}$ is lowered from $3.6\eV$ to $1.8\eV$, $0.9\eV$,
$0.4\eV$, and finally to $\epsilon_{\text{p}}=0$, where the result is
shown in Fig.~\ref{fig:ep0} (the ground state is now situated at the
$\Gamma$ point).  This is the situation for which Varma has proposed
that the current pattern sketched in Fig.~\ref{fig:pattern} would
occur.  Fig.~\ref{fig:ep0}, by contrast, shows no alignment of the
currents.

The numerical experiments for the finite cluster can, as a matter of
principle, never rule out that a symmetry is violated.  We can use
them, however, to put an upper bound on the size of the spontaneous
currents, and hence the energy associated with these currents.  If a
current pattern as sketched in Fig.~\ref{fig:pattern} were to exist,
the current correlation $\expect{j_{k,k+\hat x}j_{l,l+\hat x}}$ for
links far away from each other in a rotationally invariant ground
state should approach $\frac{1}{2}\expect{\hat x|j_{k,k+\hat x}|\hat
  x}^2$, where $\ket{\hat x}$ denotes a state with a spontaneous
current pointing in $\hat x$ direction (the factor $\frac{1}{2}$
arises because by choosing our reference link in $x$-direction, we
effectively project onto two of the four possible directions for the
current pattern).  From the values of
$10^2\!\cdot\!\expect{j_{k,k+\hat x}j_{l,l+\hat x}}$ for the four
horizontally connected links in the center of Fig.~\ref{fig:ep0},
$-0.1832, +0.0158, -0.0676$, and $+0.0158$, which should all be
positive if a current pattern were present, we estimate
$10^2\!\cdot\!\expect{j_{k,k+\hat x}}^2 < 0.20$ and hence
$\expect{\hat x|j_{k,k+\hat x}|\hat x}^2 < 4\cdot 10^{-3}$ as an upper
bound for a current pattern we are unable to detect through the
error bars of our numerical experiment.  We now denote $\expect{\hat
  x|j_{k,k+\hat x}|\hat x}$ by $j_\text{pp}$. %_\text{O-O}$.

We roughly estimate the kinetic energy $\varepsilon_\text{pp}$ per
link associated with a spontaneous current $j_\text{pp}$ of this
magnitude using $j_\text{pp}=n_\text{p}v$ and
$\varepsilon_\text{pp}=\frac{1}{2}n_\text{p} m v^2$ with
$m={1}/{2t_\text{pp}}$,
where $n_\text{p}$ is the hole density on the Oxygen sites ($n_\text{p}=0.30$
for the state of Fig.~\ref{fig:ep0}), and obtain
\begin{equation}
  \label{estimate-pp}
  \varepsilon_\text{pp}
  \approx\frac{j_\text{pp}^2}{4t_\text{pp}n_\text{p}}<5\cdot 10^{-3}.
  \nonumber
\end{equation}
A similar analysis for the CuO links, using data not shown here,
yields with $10^2\!\cdot\!\expect{j_{k,k+\hat x+\hat y}}^2 < 1.0$ and
hence $j_\text{pd}^2 < 10^{-2}$ (there is no factor
$\frac{1}{2}$ in this case)
%using $8n_\text{d}+16n_\text{p}=9$ 
an estimate of
\begin{equation}
  \label{estimate-pd}
  \varepsilon_\text{pd}
  \approx\frac{j_\text{pd}^2}{4t_\text{pd}\sqrt{n_\text{p}n_\text{d}}}
  <4 \cdot 10^{-3}.
  \nonumber
\end{equation}

We conclude that while we cannot rule out that orbital current
patterns exist, we can rule out that they are responsible for the
superconductivity, the properties of the pseudogap phase, or the
anomalous normal state properties extending up to temperatures of
several hundred Kelvin, as the energy associated with the spontaneous
loop currents is less than 5 meV per link if such currents exist.
We have assumed that the CuO planes are adequately described by the
three-band Hubbard model \eqref{3bH}, but we have allowed
$\epsilon_\text{p}$ to be much smaller than generally agreed upon, and
based our estimates on the extreme and to our purposes most
unfavorable value $\epsilon_\text{p}=0$.  (Note that the ordered
antiferromagnetic phase in undoped cuprates requires a finite
$\epsilon_\text{p}$.)  Numerical data not presented here show that our
conclusions remain intact if we set $J_\text{pd}=J_\text{pp}=0$ or/and
double the value of the repulsion $V_\text{pd}$, which generates the
orbital currents in Varma's mean-field calculation.  They also hold
for other low energy states for the finite system (\eg as situated at
the M point in the Brillouin zone for $\epsilon_\text{p}=0$).
%They are further insenistive to changing the primitive region for the
%PBCs from being spanned by the vectors $(2,2$ and $(2,-2)$ to being
%spanned by $(3,1)$ and $(2,-2)$.

Nonetheless, we should keep in mind that any analysis of a model can
only reach a conclusion valid for this model.  The question of whether
current patterns exist in CuO superconductors can ultimately only be
settled by experiment.  We consider it likely, however, that an
eventual consensus among experiments will confirm our conclusion.

%A more elaborate account of this work is in preparation.
\begin{acknowledgments}
  We wish like to thank C.M.\ Varma, V.\ Aji, F.\ Evers and in
  particular P.\ W\"olfle for many illuminating discussions of this
  subject.  RT was supported by a PhD scholarship from the
  Studienstiftung des deutschen Volkes.  We further acknowledge the
  support of the computing facilities of the INT at the
  Forschungszentrum Karlsruhe.
\end{acknowledgments}

% Create the reference section using BibTeX:
%\bibliographystyle{/users/tkm/rachel/bib/prsty}
%\bibliography{/users/tkm/rachel/bib/book,/users/tkm/rachel/bib/paper,/users/tkm/rachel/bib/unpub,/users/tkm/rachel/bib/htc}

\end{document}